\documentclass[aps,preprint]{revtex4}%
\usepackage{amsfonts}
\usepackage{amsmath}
\usepackage{amssymb}
\usepackage{graphicx}%
\begin{document}
\title{Sub-Doppler spectroscopy of Rb atoms in a sub-micron vapor cell in the
presence of a magnetic field}
\author{David Sarkisyan, Aram Papoyan, Tigran Varzhapetyan}
\affiliation{Institute for Physical Research, NAS of Armenia, Ashtarak-2 378410 Armenia}
\author{Janis Alnis, Kaspars Blush, Marcis Auzinsh}
\affiliation{Department of Physics, University of Latvia, 19 Rainis blvd, Riga, LV-1585, Latvia}
\date{\today}

\begin{abstract}
We report the first use of an extremely thin vapor cell (thickness $\sim400$
nm)\ to study the magnetic-field dependence of laser-induced-fluorescence
excitation spectra of alkali atoms. \ This thin cell allows for sub-Doppler
resolution without the complexity of atomic beam or laser cooling techniques.
\ This technique is used to study the laser-induced-fluorescence excitation
spectra of Rb in a $50$ G magnetic field. \ At this field strength the
electronic angular momentum $\mathbf{J}$ and nuclear angular momentum
$\mathbf{I}$ are only partially decoupled. \ As a result of the mixing of
wavefunctions of different hyperfine states, we observe a nonlinear Zeeman
effect for each sublevel, a substantial modification of the transition
probabilities between different magnetic sublevels, and the appearance of
transitions that are strictly forbidden in the absence of the magnetic field.
\ \ For the case of right- and left- handed circularly polarized laser
excitation, the fluorescence spectra differs qualitatively. Well pronounced
magnetic field induced circular dichroism is observed. \ These observations
are explained with a standard approach that describes the partial decoupling
of $\mathbf{I}$ and $\mathbf{J}$ states.

\end{abstract}
\maketitle

\section{ Introduction}

Sub-Doppler probes of atoms and molecules in a thermal gas cell is a
well-established subfield of laser spectroscopy. In particular,
velocity-selective techniques such as saturated absorption spectroscopy,
polarization spectroscopy, Doppler-free two photon spectroscopy
\cite{sch1,let1}, coherent dark state spectroscopy \cite{wyn1} and other
techniques can provide linewidths orders of magnitude smaller than the Doppler width.

During the past several years a fundamentally different approach has been
developed. \ This new method is based on the spectroscopy of a diluted atomic
vapor in an extremely thin cell. \ In most of these studies thermal Cs atoms
are confined to cells of thickness of $10$ $\mu$m to $1000$ $\mu$m. \ The Cs
atoms that are moving parallel to the cell walls are laser excited and
fluoresce in much the same way as atoms in a normal vapor cell. \ Atoms moving
perpendicularly to the closely-spaced cell walls are likely to suffer
collisions with the wall before they absorb and fluoresce and therefore do not
make a significant contribution to the laser-induced fluorescence signal.
\ Thus the spectra obtained is that of atoms that show no first-order Doppler
shift. \ This approach allows for a relatively simple experimental apparatus
to provide sub-Doppler resolution \cite{gla1,ai1,bri1,bri2,bri3,bri4,izm1}.

It was demonstrated a long time ago in microwave spectroscopy \cite{rom1}, and
recently in the optical domain \cite{var1,zam1,yar1,dut1}, that the optimal
cell thickness\ $L$ for atomic line shape narrowing is approximately
$\lambda/2$ where $\lambda$ is the wavelength of spectral features to be
observed. \ This fact stimulated us (D.S. and A.P.) to develop an extremely
thin gas cell of thickness of $100-300$ nm. \ As far as we know this is an
absolute record. The cell used in the present investigation is at least one
order of magnitude thinner than all other cells used in this manner and has
allowed for the first optical measurement of atoms in a cell which satisfies
the condition $L\approx\lambda/2$. These conditions allowed recently to
resolve all hyperfine transitions of the D$_{2}$ line of Cs atoms. In addition
to the Cs study, a naturally occurring isotopic mixture of Rb atoms was
introduced to the thin cell. For this case, $^{85}$Rb and $^{87}$Rb hyperfine
transitions were well resolved in the first D$_{1}$ line fluorescence spectrum
\cite{sar2}. \ In addition to removing Doppler broadening, thin cells provide
a second advantage that is not evident in other types of spectroscopy: \ In an
ordinary vapor cell it is extremely easy to reach laser intensities that
change the atomic ground state magnetic sublevel population (optical pumping)
and to saturate transitions. In ordinary cells optical pumping of atomic
states typically occurs at laser radiation intensities of $10-1000$
nW/cm$^{2}$ complicating the experimental determination of relative absorption
strengths and/or atomic populations. \ In a thin cell, however, the population
of atoms that do not move parallel to the cell walls constantly replenish
depleted ground-state population, leading to linear laser absorption at laser
intensities up to at least, 50 mW/cm$^{2}$ \cite{sar2}.

In the present paper we demonstrate that a sub-micron cell not only allows for
sub-Doppler atomic spectroscopy, but more sophisticated studies of linear
magneto-optical effects as well. \ Specifically, we use sub-micron
cells\ filled with $^{85}$Rb and $^{87}$Rb to explore circular dichroism
induced by an external magnetic field. The effect of magnetic field-induced
circular dichroism is important for a variety of reasons. For some time there
have been attempts to use polarization plane rotation of a laser beam by an
atomic or molecular gas to test of CP parity violation caused by interference
between the neutral weak and electromagnetic interactions that couple a
valence electron to the nucleus \cite{khr1,com1,bou1,gue1}. Alkali atoms
including Cs are considered as a perspective objects for such CP parity
violation measurements. Simple linear magneto optical effects caused by stray
magnetic fields in general and magnetic field induced circular dichroism in
particular can mimic the effect of rotation of the laser beam polarization
plane by parity-violating neutral currents. \ It is therefore essential that
the less exotic linear magneto-optical effect be completely understood. \ In a
more practical application, magnetic field induced circular dichroism in
alkali atoms has been successfully applied for a diode laser frequency
stabilization based on difference of the absorption of right-hand and
left-hand circular polarized radiation in atomic vapor placed in the constant
magnetic field \cite{yas1,cli1}. This technique was applied to diode lasers in
the work of the Joint Institute for Laboratory Astrophysics (JILA) group
\cite{cor1} who coined for this stabilization method the term\textit{ dichroic
atomic vapor laser lock} (DAVLL).

In general circular dichroism studies in alkali metal vapors have a long
history. The first detailed study of magneto optical rotational effects for
closely lying hyperfine levels were performed in connection with the
non-conservation of CP parity in bismuth \cite{nov1,rob1}. The work of G.J.
Roberts et al is a combined theoretical and experimental investigation of
magneto-optical effects on Doppler broadened transitions \cite{rob1}. \ The
primary focus of this work is Faraday rotation but the closely related
phenomenon of magnetic field induced circular dichroism is discussed as well.
\ The shift of the energy of magnetic sublevels in the magnetic field is
considered and hyperfine level mixing caused by the field is taken into
account as well. Doppler broadened optical rotation spectra and, in one
experiment, circular dichroism spectra are observed. Typical features of\ the
spectra have widths of the order of GHz. \ In a more recent paper \cite{che1},
Doppler broadened magneto-optical rotation spectra in a room-temperature Cs
vapor in a magnetic field of the order of $45$ G, are presented, along with a
corresponding theoretical treatment. \ An interesting result of the
theoretical treatment is the fact that even at this relatively weak magnetic
field, the perturbation of the wave function plays an essential role in the
Zeeman energy level shift. In a study carried out in Quebec \cite{tre1} both
isotopes of rubidium $^{85}$Rb and $^{87}$Rb as well as Cs atoms are used to
study absorption spectra of atoms in presence of a magnetic field. In this
study relatively strong magnetic field strengths of $790$, $1440$ and $2300$ G
are applied to the gas cell. It is shown that a magnetic field strength
exceeding $1000$ G is required to observe magnetic structure in\ the Doppler
broadened absorption spectra. At this high field values hyperfine coupling is
broken and nuclear and electronic moments in atom are decouple. \ Studies of
linear magneto-optical methods can be carried out also by non-direct methods
like selective reflection spectroscopy \cite{wei1,pap1} but will not be
discussed in this work.

In present paper we demonstrate that extremely thin gas cells can be used
successfully to study with sub-Doppler resolution magnetic field induced
circular dichroism in atoms. We are not aware of any previous experimental
study of magnetic field induced circular dichroism that employes a sub-Doppler
technique, although a theoretical study of magneto-optical effects on atoms in
a thin cell has been carried out previously \cite{zam1}. Here we observe
magnetic field induced circular dichroism. \ We measure the total fluorescence
intensity as dependent on the scanned diode laser frequency
(leser-induced-fluorescence excitation spectra) for both Rb isotopes.
Excitation with circularly polarized as well as linearly polarized radiation
is used. It is demonstrated that the laser-induced fluorescence excitation
spectra is qualitatively very different for left and right circularly
polarized radiation. In contrast, the corresponding Doppler broadened spectra
changes only slightly as the light helicity changes and this change may be
explained with a simple model that ignores the rich physics observed in the
sub-Doppler spectra.

The paper is organized as follows. In Section II the experimental procedure
and obtained results are presented for the two naturally occurring Rb
isotopes. Section III compares the observed spectra with model calculations.
In Section IV results are discussed, and in Section V conclusions are drawn.

\section{Experimental}

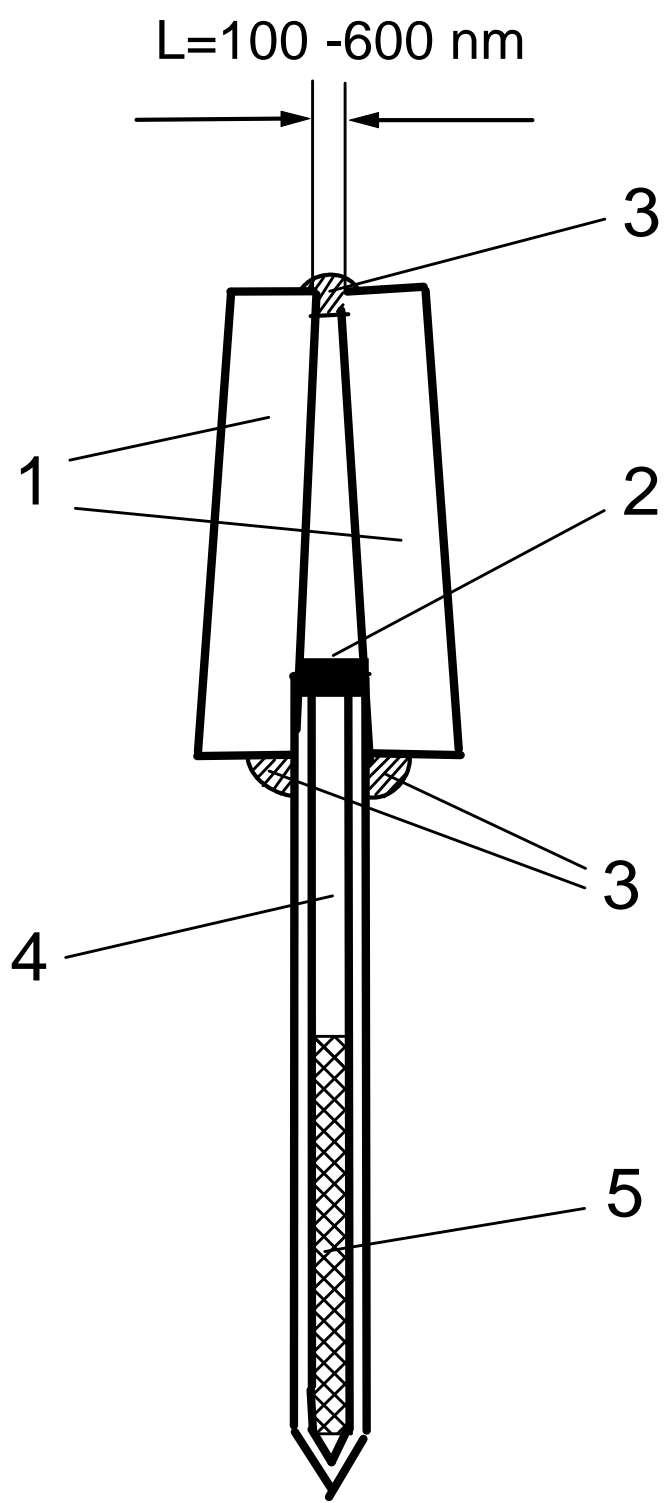
\begin{figure}
[h]
\begin{center}
\phantom{\rule{2.7994in}{0.4454in}}\caption{The design of the ETC. $1$ -- the
windows of the ETC, made from a garnet (YAG) crystal, the thickness is $3$ mm,
diameter is $30$ mm. The flatness of the both inner surfaces of the YAG
windows is less than $\lambda/10$; $2$ -- $\sim$\ $600$ nm-thick Al$_{2}%
$O$_{3}$\ coating in the shape of a $1$ mm-wide and $10$ mm-long strip; $3$ --
the glue; $4$ -- the sapphire side arm with a \textquotedblright molybdenum
glass\textquotedblright\ termination; $5$ -- column of Rb metal.}%
\label{fig1}%
\end{center}
\end{figure}

The design of the ETC is presented in Fig. \ref{fig1}. It is a new
modification of sapphire cells developed earlier (see \cite{sar2} and
references therein). The windows of 30 mm diameter and 3 mm thickness are made
from very well-polished garnet (YAG) crystal, which is resistant to highly
corrosive alkaline vapors. The roughness of the inner surfaces of the YAG
windows is $<\lambda/10$. In order to realize a wedged gap between the YAG
windows, a $1$ mm-wide and $10$-mm-long Al$_{2}$O$_{3}$\ strip coating of
$\approx$\ $600$ nm thickness has been deposited on the surface of one of the
YAG windows, near the lower edge. The sapphire side arm with a molybdenum
glass termination is used for filling the cell with Rb in the same way as for
all-glass cells. To attach the side arm, a hole is drilled at the bottom of
the YAG windows. Before filling the ETC with natural mixture of rubidium atoms
($72.15$\% of $^{85}$Rb with nuclear spin $I=5/2$ and $27.85$\% of $^{87}$Rb
with nuclear spin $I=3/2$), the cell was carefully out-gassed.

The wedged gap between the windows allows us to exploit a $\sim$\ $\lambda/2$
thickness of a vapor layer that is important for having the narrowest
sub-Doppler line width \cite{yar1}.\ An interferometric mapping of the gap
thickness is performed using He-Ne laser. This measurement shows that the
thickness of the gap varies in the range of $100$ nm - $600$ nm. A $\sim
$\ $400$ nm thickness region of the cell aperture ($\approx$ $\lambda/2$) has
been explored in present experiment.

The upper temperature limit for the ETC is $\sim$\ $400$ $%
%TCIMACRO{\U{b0}}%
%BeginExpansion
{{}^\circ}%
%EndExpansion
$C. The ETC operated with a specially designed oven, which has four openings;
two for the laser beam transmission and two that allow for fluorescence to be
detected simultaneously from two different directions. \ The density of the Rb
atomic vapor is determined by the temperature of the boundary of Rb metal
column inside the side arm (Fig \ref{fig1}). The typical value of metal
boundary (side arm) temperature is kept at $120$ $%
%TCIMACRO{\U{b0}}%
%BeginExpansion
{{}^\circ}%
%EndExpansion
$C throughout the measurements, while the window temperature is set to
somewhat higher value ($140$ $%
%TCIMACRO{\U{b0}}%
%BeginExpansion
{{}^\circ}%
%EndExpansion
$C) to prevent Rb condensation. \ The corresponding number density of Rb atoms
is $N_{Rb}\approx2\times10^{13}$\ cm$^{-3}$, and the collisional broadening is
$\sim3$ MHz. The Doppler line width in an ordinary cell at this temperature
regime is $\approx$ $600$ MHz.

The ETC with Rb atoms is placed in $3$ pairs of mutually perpendicular
Helmholtz coils which cancel the ambient magnetic field and provide a
homogeneous magnetic field in an arbitrary direction. The geometry of the
experiment is depicted schematically in Fig. \ref{geom}. The collimated
radiation beam of the cw laser diode ($\lambda$ $=780.2$ nm, $25$ MHz line
width, single mode, $\oslash$ $3$ mm.) is directed at normal incidence into
the Rb-filled $\sim400$-nm-thick region of the ETC. The maximum laser
intensity is $50$ mW/cm$^{2}$. A\ Glan-Thomson prism is used to purify the
linear radiation polarization of the laser. \ To produce a circular
polarization, a $\lambda/4$ plate is utilized. \ A photodiode\textsc{\ }with
an aperture of $1$ cm$^{2}$ is placed at $90%
%TCIMACRO{\U{b0}}%
%BeginExpansion
{{}^\circ}%
%EndExpansion
$ to the laser propagation direction to detect the fluorescence signal
emerging through one of the two side openings of the cell oven. The photodiode
collects emission within $\sim0.1$ srad solid angle. The signal of the
photodiode is amplified and recorded with a storage oscilloscope. \ The
intensity of the fluorescence emission (with no spectral and polarization
resolution) from the vapor layer excited by linearly and circularly polarized
radiation is recorded versus the laser radiation frequency. To obtain the
laser-induced-fluorescence excitation spectra, the laser frequency is scanned
linearly in a\ $9$ GHz spectral region around the Rb D$_{2}$ line by means of
injection current ramping.

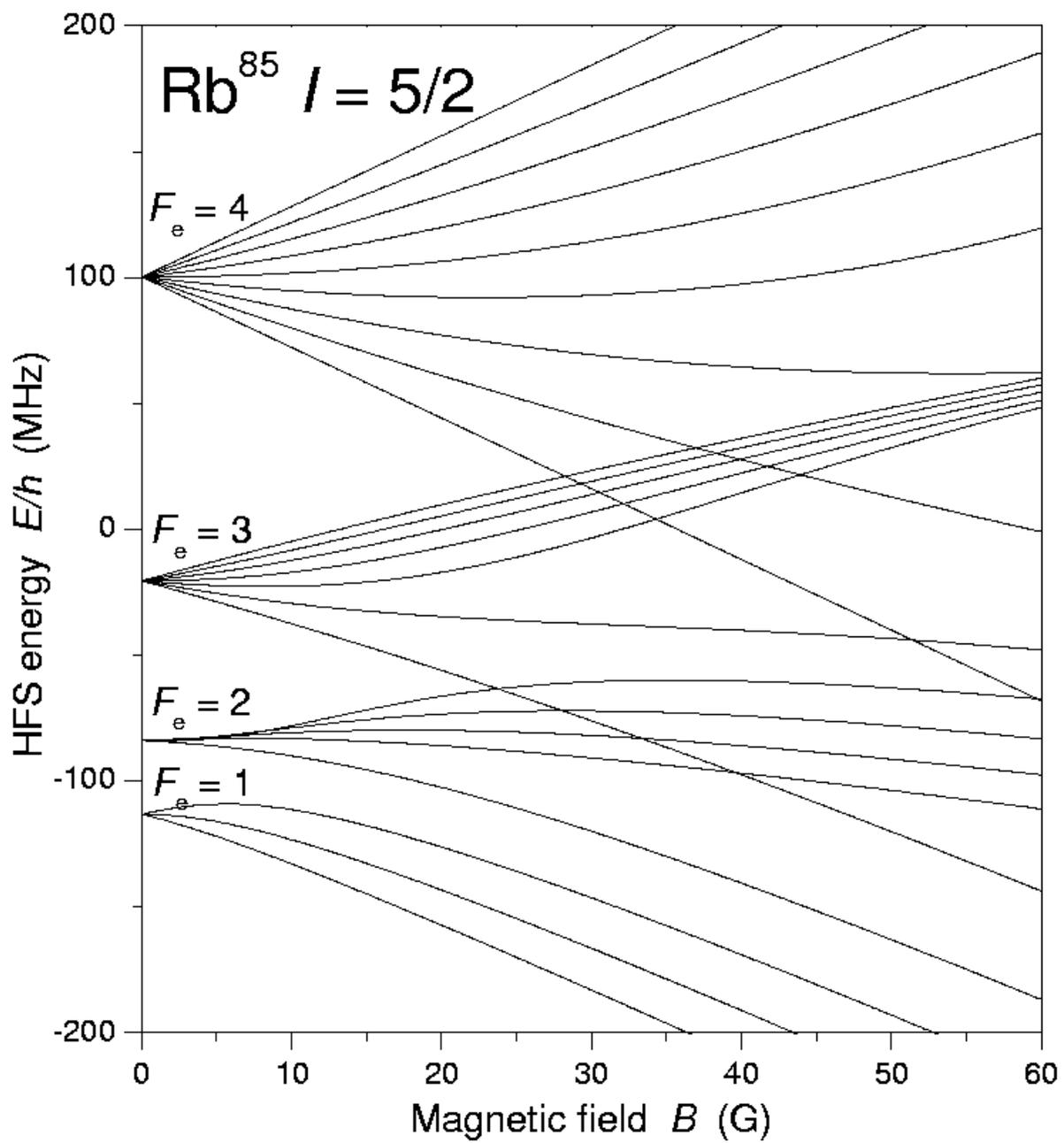
\begin{figure}
[hh]
\begin{center}
\phantom{\rule{2.7994in}{0.4454in}}\caption{Excited-state hyperfine magnetic
sublevel splitting for intermediate strength of a magnetic field for $^{85}$Rb
isotope}%
\label{Rb85energy}%
\end{center}
\end{figure}

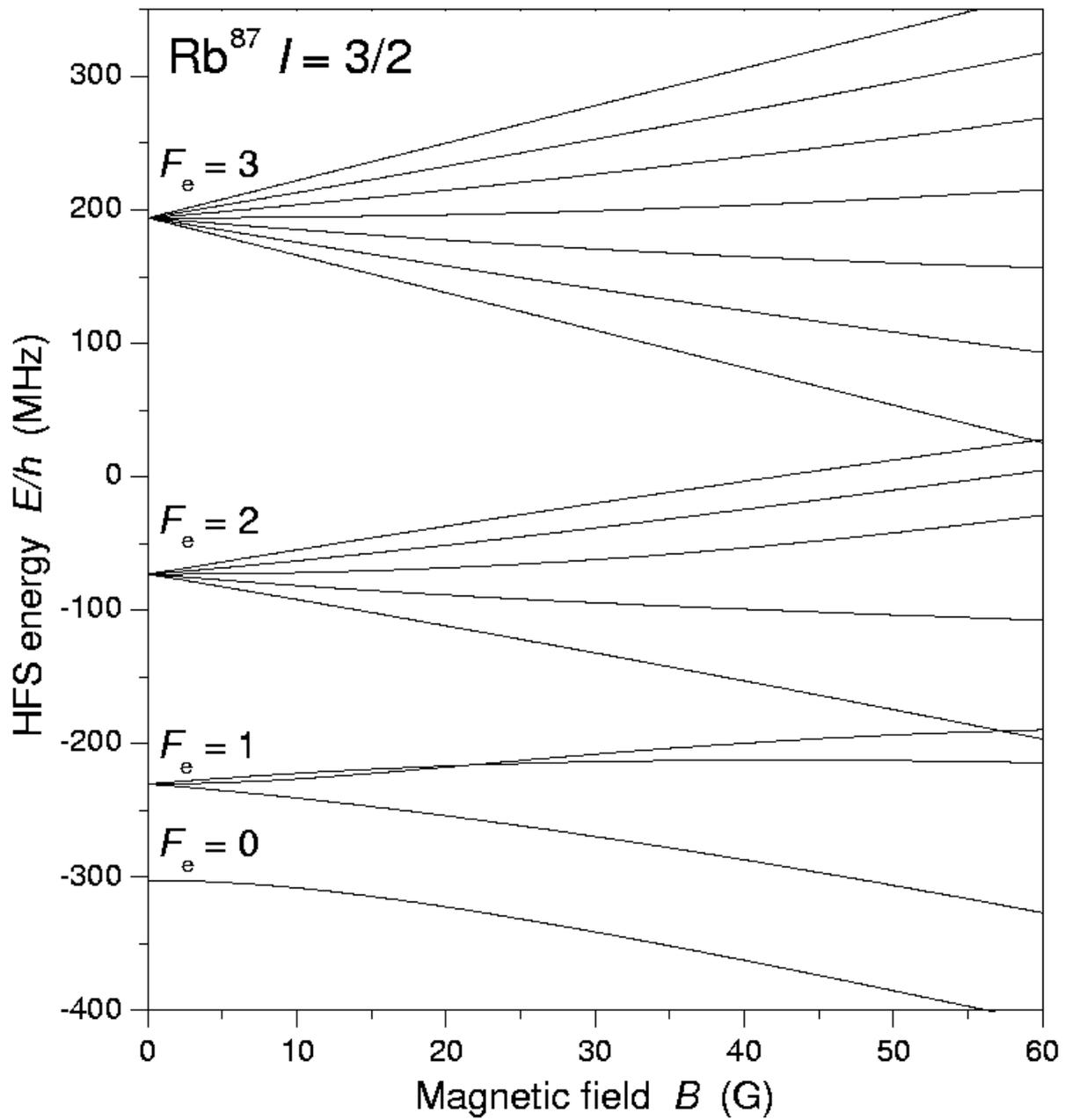
\begin{figure}
[hhh]
\begin{center}
\phantom{\rule{2.7994in}{0.4454in}}\caption{Excited-state hyperfine magnetic
sublevel splitting for intermediate strength of a magnetic field for $^{87}$Rb
isotope.}%
\label{Rb87energy}%
\end{center}
\end{figure}

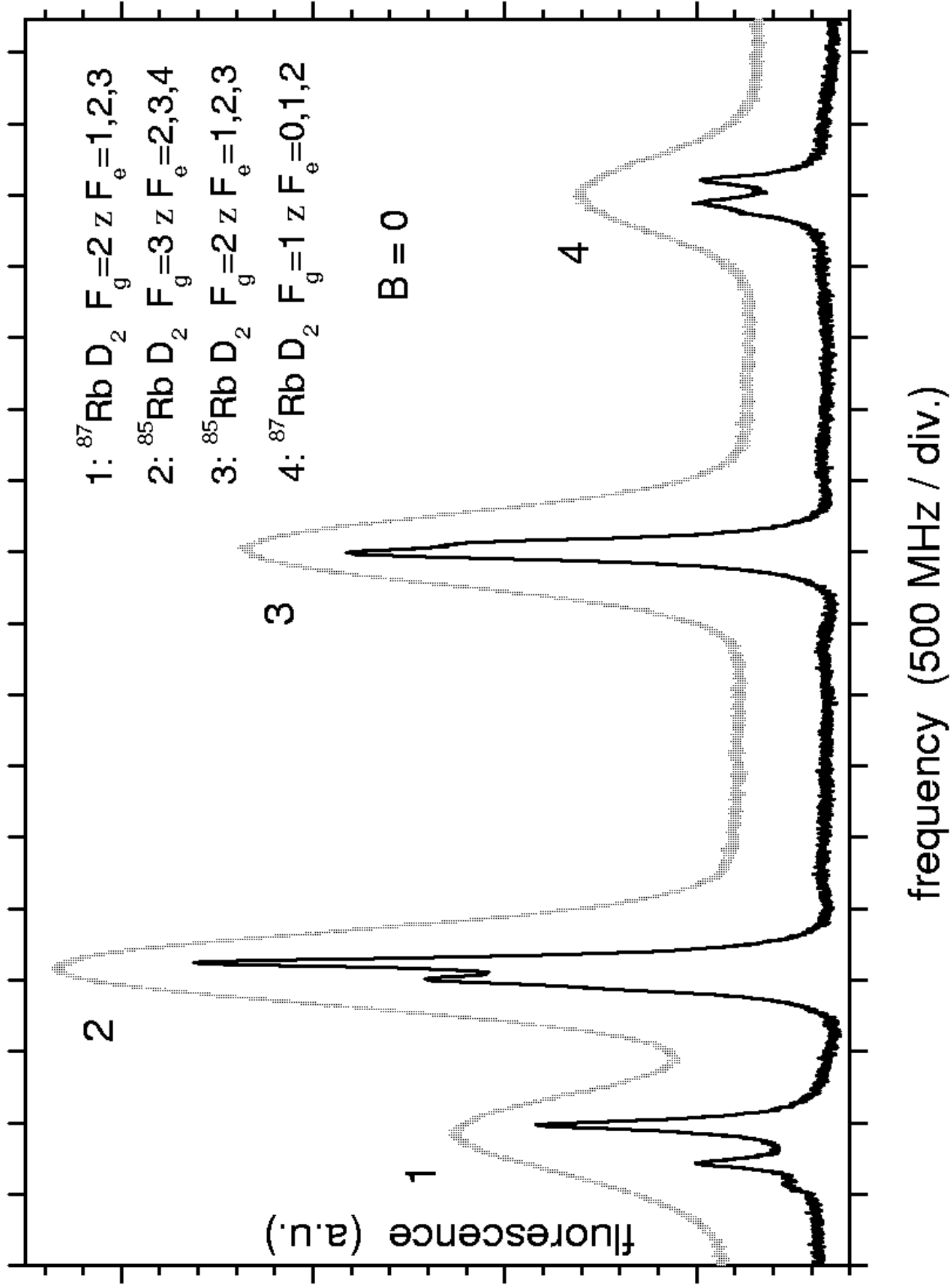
\begin{figure}
[hhhh]
\begin{center}
\phantom{\rule{2.7994in}{0.4454in}}\caption{Doppler broadened fluorescence
spectra (light line) obtained in an ordinary cell and sub-Doppler spectra
obtained in ETC on D$_{2}$ line of both Rb isotopes in absence of the magnetic
field.}%
\label{spectra_total}%
\end{center}
\end{figure}

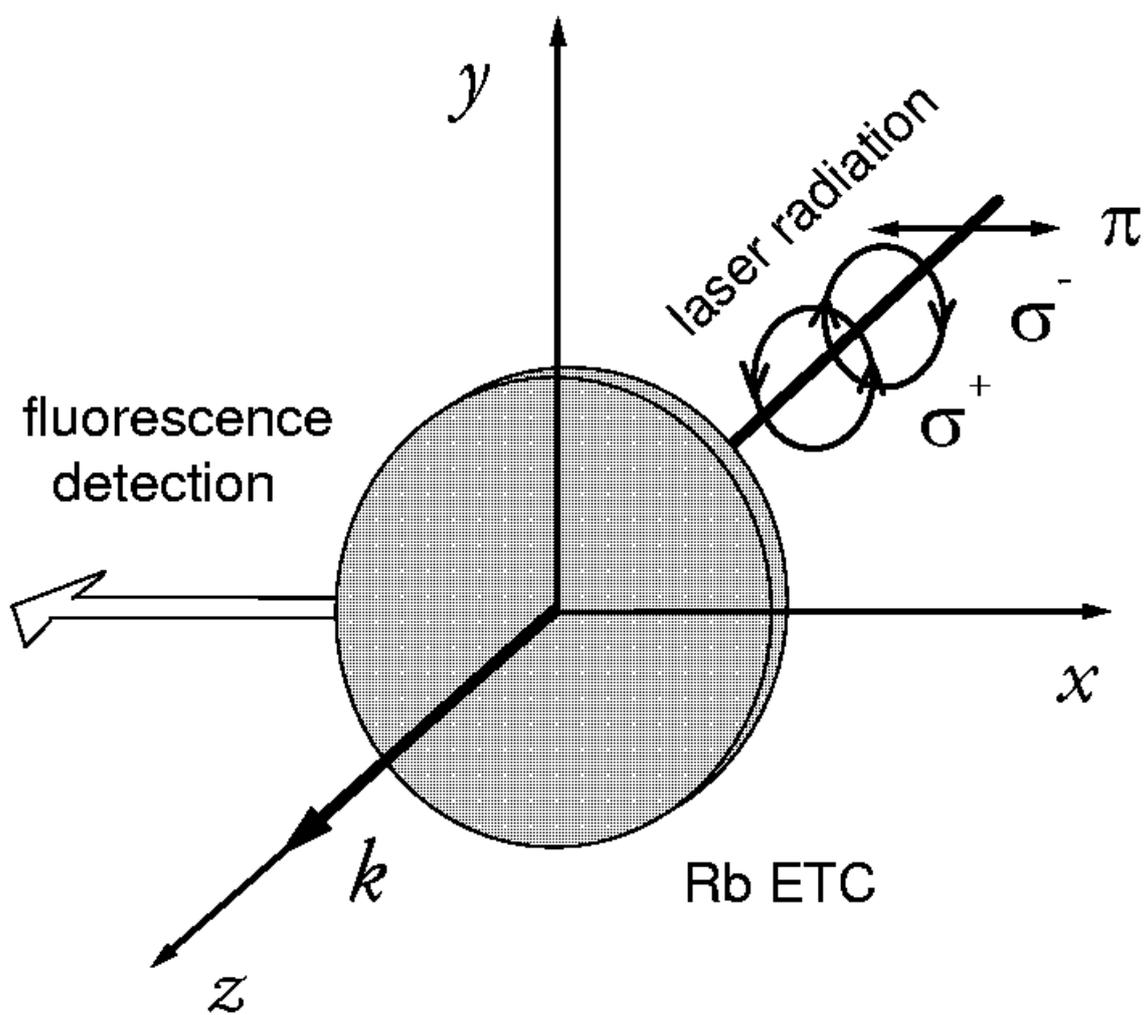
\begin{figure}
[hhhhh]
\begin{center}
\phantom{\rule{2.7994in}{0.4454in}}\caption{Geometrical configuration of the
experiment. Magnetic field is applied along $z$ or $x$ axis. }%
\label{geom}%
\end{center}
\end{figure}

As a result of hyperfine interactions, the ground state level of $^{85}$Rb is
split into two components with total angular momentum quantum numbers
$F_{g}=2$ and $F_{g}=3$ and the ground-state level of $^{87}$Rb is split into
two components with total angular momentum quantum numbers $F_{g}=1$ and
$F_{g}=2$. The ground state level splitting for $^{85}$Rb and $^{87}$Rb are
approximately $3$ and $7$ GHz, respectively. In contrast, the four
excited-state hyperfine components are separated by only several hundred MHz
(see Figs. \ref{Rb85energy}\ and \ref{Rb87energy}). Since the separation of
the hyperfine sub-levels of the $5P_{3/2}$\ state with $F_{e}=0-3$ for $^{87}%
$Rb ($72-157-267$ MHz) and $F_{e}=1-4$ for $^{85}$Rb ($29-63-121$ MHz) is less
than the Doppler broadening ($\sim$\ $600$ MHz), the D$_{2}$\ spectrum for an
ordinary Rb cell consists of only two resolved lines for each Rb isotope:
$5S_{1/2}\ (F=1)\rightarrow\ 5P_{3/2}\ (F=0-2)$ and $5S_{1/2}%
\ (F=2)\rightarrow\ 5P_{3/2}\ (F=1-3)$ for $^{87}$Rb and $5S_{1/2}%
\ (F=2)\rightarrow\ 5P_{3/2}\ (F=1-3)$ and $5S_{1/2}\ (F=3)\rightarrow
\ 5P_{3/2}\ (F=2-4)$ for $^{85}$Rb. The total laser-induced-fluorescence
excitation spectra on a full D$_{2}$\ line scan in the absence of the magnetic
field are presented in figure \ref{spectra_total}. The gray curve shows the
Doppler broadened spectrum in an ordinary room-temperature cm-long cell. The
black curve shows the same spectrum in an ETC. In the latter, features
representing hyperfine levels of the excited state can be seen clearly.

To explore magneto-optical effects, we register the fluorescence spectra for
three different polarizations of laser radiation. Namely, left- and right-
handed polarization for the case of laser propagation parallel to the magnetic
field direction ($\sigma^{+}$ and $\sigma^{-}$) and linear polarization along
the magnetic field direction ($\pi$). Our choice of laser radiation
polarization and magnetic field direction is determined by the fact that for
each of these geometries, only one well defined transition type between
magnetic sublevels of ground state $m_{F_{g}}$ and magnetic sublevels of
excited state $m_{F_{e}}$ can take place. Namely\textsc{\ }$\Delta m=m_{F_{e}%
}-m_{F_{g}}$\textsc{\ }is equal to $+1$ for $\sigma^{+}$ excitation, is equal
to $-1$ for $\sigma^{-}$ excitation, and is equal to $0$ for $\pi$ excitation.
We record experimental laser-induced-fluorescence excitation spectra of Rb
atoms at a $B$-field strength of $50$ G. \ This intermediate field strength is
chosen so that the Pashen - Back effect is important. At this interesting
field strength, the magnetic field starts to decouple the nuclear spin
$\mathbf{I}$\textbf{\ }from the electronic angular momentum $\mathbf{J}$ but
these moments still remain partially coupled. \ As a rule at this field
strength crossing take place in the magnetically shifted hyperfine levels (see
Figs. \ref{Rb85energy} and \ref{Rb87energy}.)

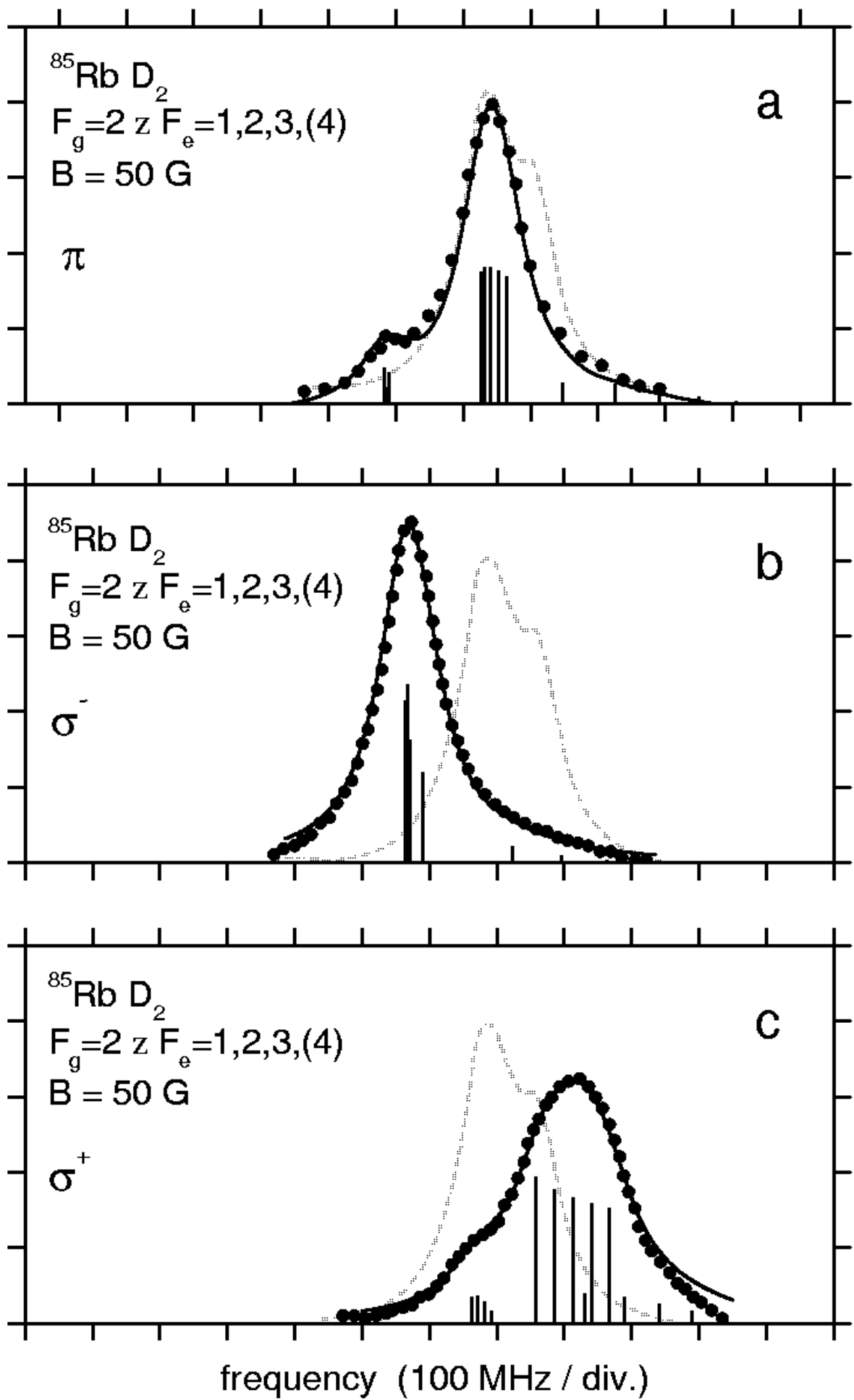
\begin{figure}
[hhhhhh]
\begin{center}
\phantom{\rule{2.7994in}{0.4454in}}\caption{Experimental (scatter circle) and
simulated (dark line) fluorescence spectra of $^{85}$Rb isotope on transitions
$F_{g}=2\rightarrow F_{e}=1,2,3,(4)$. ($a$) linearly polarized $\pi$
excitation, ($b$) circularly polarized $\sigma^{-}$ excitation, ($c$)
circularly polarized $\sigma^{+}$ excitation. With light line respective
fluorescence spectra in absence of the magnetic field is shown.}%
\label{exp1}%
\end{center}
\end{figure}

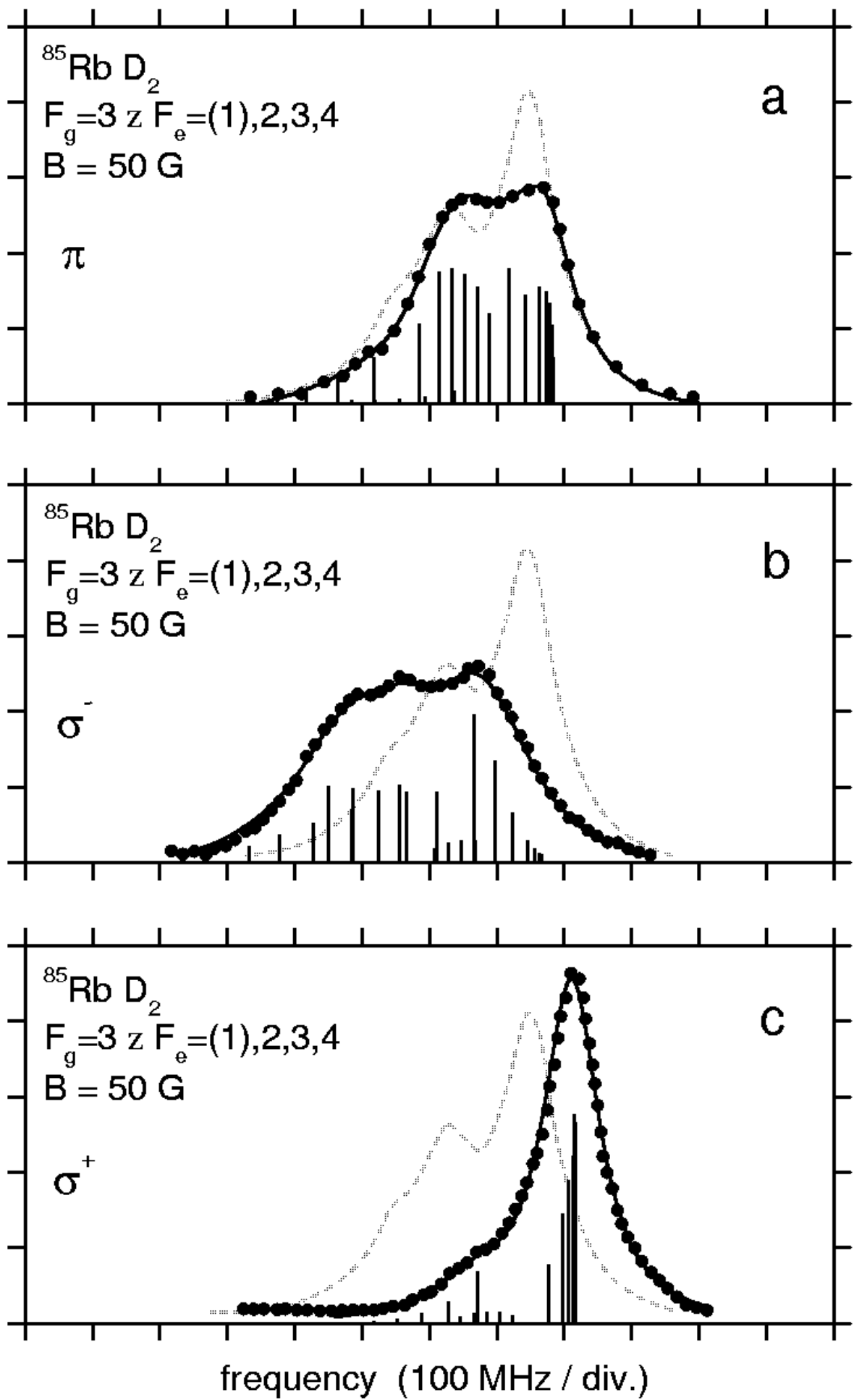
\begin{figure}
[hhhhhhh]
\begin{center}
\phantom{\rule{2.7994in}{0.4454in}}\caption{Experimental (scatter circle) and
simulated (dark line) fluorescence spectra of $^{85}$Rb isotope on transitions
$F_{g}=3\rightarrow F_{e}=(1),2,3,4$. ($a$) linearly polarized $\pi$
excitation, ($b$) circularly polarized $\sigma^{-}$ excitation, ($c$)
circularly polarized $\sigma^{+}$ excitation. With light line respective
fluorescence spectra in absence of the magnetic field is shown.}%
\label{exp2}%
\end{center}
\end{figure}

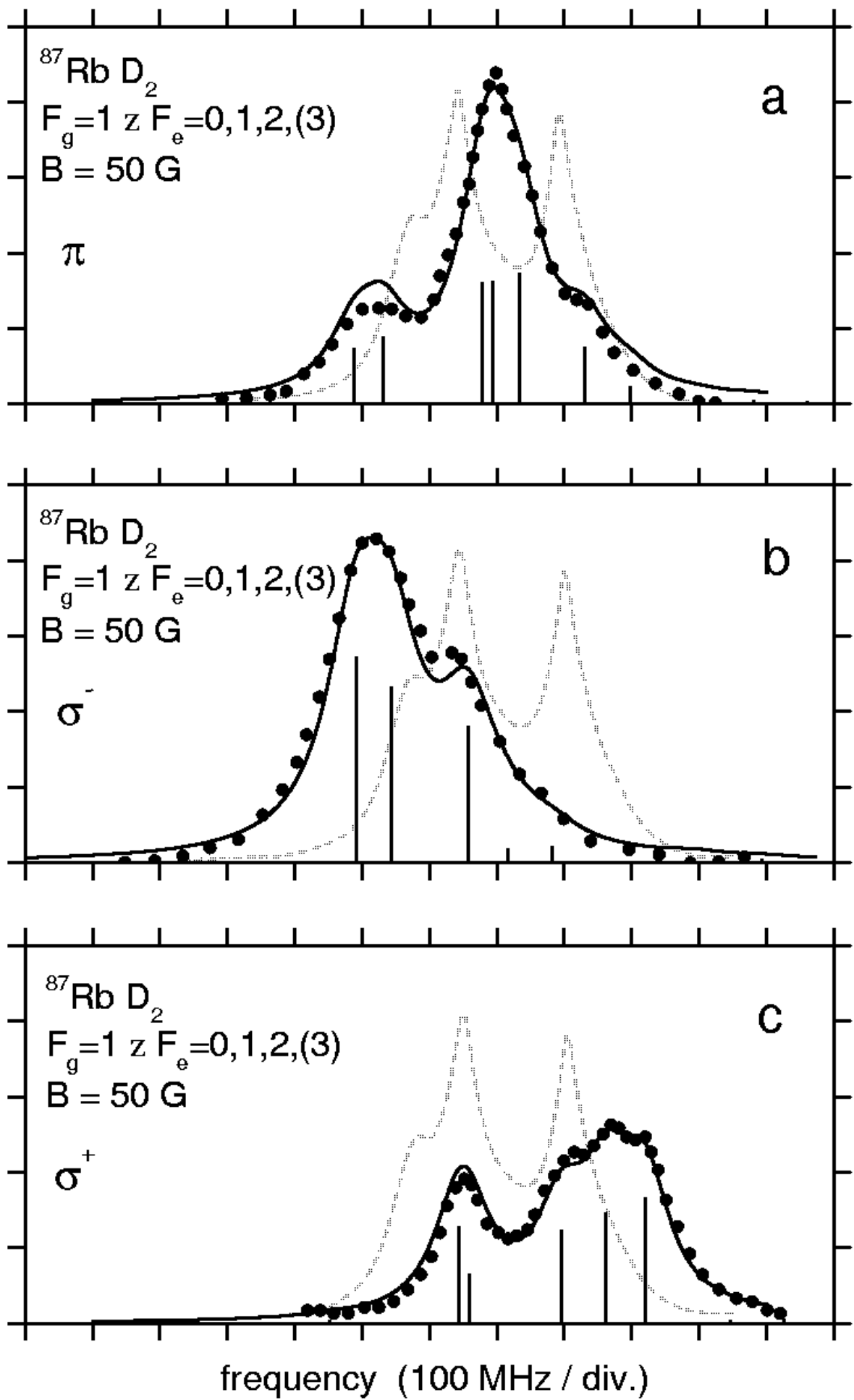
\begin{figure}
[hhhhhhhh]
\begin{center}
\phantom{\rule{2.7994in}{0.4454in}}\caption{Experimental (scatter circle) and
simulated (dark line) fluorescence spectra of $^{87}$Rb isotope on transitions
$F_{g}=1\rightarrow F_{e}=0,1,2,(3)$. ($a$) linearly polarized $\pi$
excitation, ($b$) circularly polarized $\sigma^{-}$ excitation, ($c$)
circularly polarized $\sigma^{+}$ excitation. With light line respective
fluorescence spectra in absence of the magnetic field is shown.}%
\label{exp3}%
\end{center}
\end{figure}

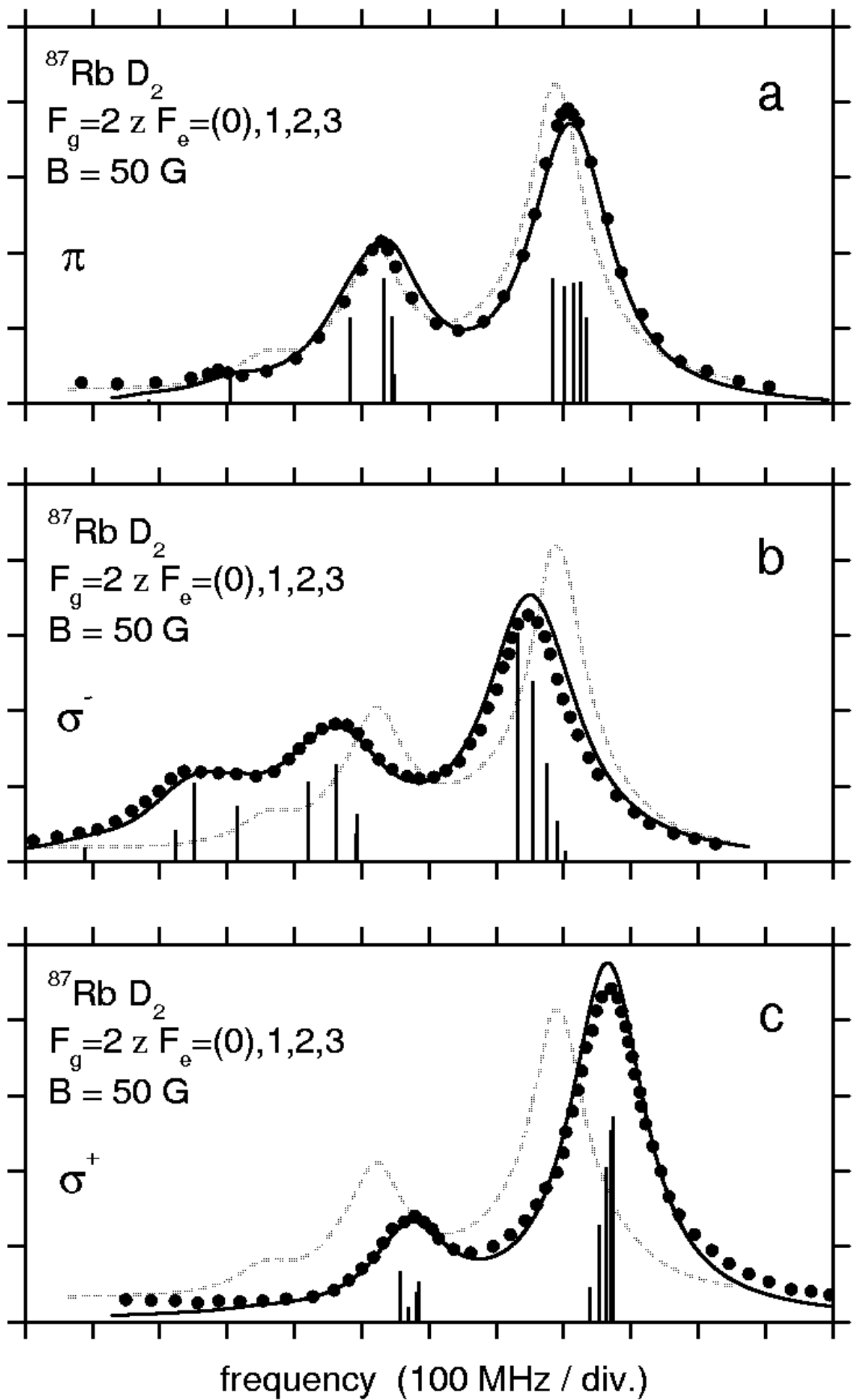
\begin{figure}
[hhhhhhhhh]
\begin{center}
\phantom{\rule{2.7994in}{0.4454in}}\caption{Experimental (scatter circle) and
simulated (dark line) fluorescence spectra of $^{87}$Rb isotope on transitions
$F_{g}=2\rightarrow F_{e}=1,2,3,(4)$. ($a$) linearly polarized $\pi$
excitation, ($b$) circularly polarized $\sigma^{-}$ excitation, ($c$)
circularly polarized $\sigma^{+}$ excitation. With light line respective
fluorescence spectra in absence of the magnetic field is shown.}%
\label{exp4}%
\end{center}
\end{figure}

In Figures \ref{exp1}, \ref{exp2}, \ref{exp3}, \ref{exp4} scattered circles
depict experimentally obtained signals for the following transitions $^{85}%
$Rb, absorption $F_{g}=2\rightarrow F_{e}=1,2,3,(4)$; $^{85}$Rb, absorption
$F_{g}=3\rightarrow F_{e}=(1),2,3,4$; $^{87}$Rb, absorption $F_{g}%
=1\rightarrow F_{e}=0,1,2,(3)$ and $^{87}$Rb, absorption $F_{g}=2\rightarrow
F_{e}=(0),1,2,3$. \ For reference in the same figures with lighter color are
depicted experimentally measured fluorescence spectra in absence of the
magnetic field. \ We note the observation of transitions that violate the
$\Delta F=0,\pm1$ selection rule for electric dipole transitions. This
selection rule is strictly true only for hyperfine levels in absence of the
magnetic field. As the magnetic field intensity increases from zero, the
hyperfine levels mix and transitions that are forbidden in absence of the
field start to be allowed. \ The excited-state quantum number for transitions
that are not allowed in the absence of a magnetic field are written in
parenthesis. \ 

From the figures presented it is clearly seen that laser-induced-fluorescence
excitation spectrum for left and right-handed excitation are qualitatively
different. From simple symmetry considerations, these spectra must coincide in
absence of the magnetic field. In presence of the field helicity of the laser
polarization with respect to the field direction starts to play a very
important role and these spectra are very different for all transitions
presented. \ In agreement with nearly linear fluorescence intensity dependence
when the intensity of the diode laser is increased, reported in \cite{sar2},
also in case of Rb the shape of fluorescence spectrum does not noticeably
depend on laser radiation intensity, at least in the range of $I_{L}=$ $1-50$
mW/cm$^{2}$. All the spectra presented in graphs were recorded with $I_{L}=$
$50$ mW/cm$^{2}$.

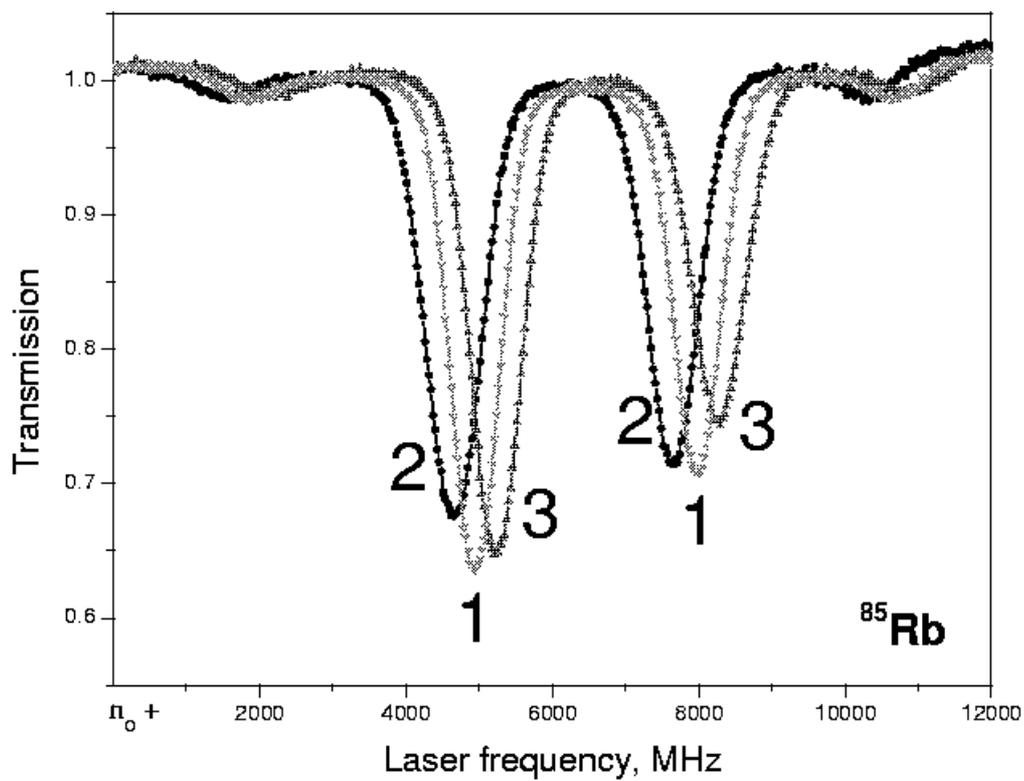
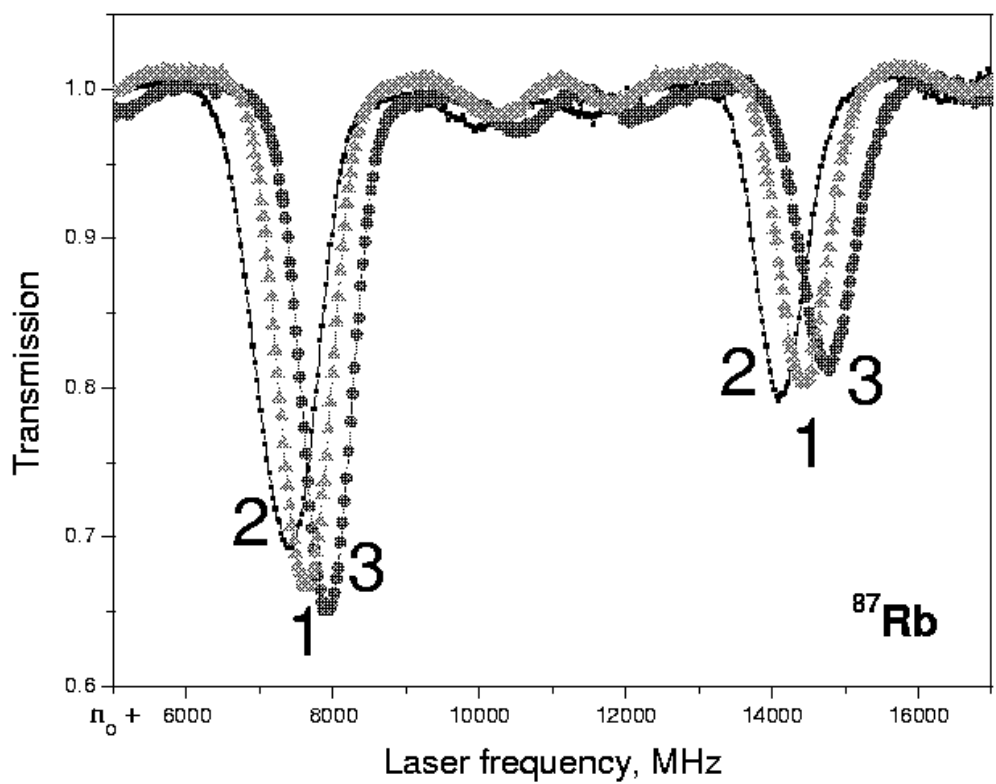
\begin{figure}
[hhhhhhhhhh]
\begin{center}
\phantom{\rule{2.7994in}{0.4454in}}\caption{Absorption spectra in an ordinary
optical cell for ($a$) $^{85}$Rb and ($b$) $^{87}$Rb. Peak 1 is absorption for
$\sigma^{-}$ or $\sigma^{+}$ radiation in absence of the magnetic field. Peak
2 corresponds to the absorption of $\sigma^{-}$ radiation in a magnetic field
and peak 3 to the absorption of $\sigma^{+}$ radiation.}%
\label{exp_broad}%
\end{center}
\end{figure}

It is interesting to compare these spectra for case of ordinary cells when
only Doppler broadened spectra can be recorded. In figure \ref{exp_broad}
absorption spectra obtained at room temperature in an ordinary call containing
pure isotopes of Rb atoms. In one case $^{85}$Rb and in other case $^{87}$Rb.
Unfortunately, for technical reasons only the absorption and not the
laser-induced-fluorescence excitation spectra were taken. Nevertheless, in
these cells absorption and laser-induced-fluorescence excitation spectra
should be very similar. We see that in these Doppler broadened spectra, both
isotopes lead to spectra that are qualitatively similar for both circularities
of laser radiation. \ Each isotope leads to only one peak for absorption from
each of ground state hyperfine levels. This peak is shifted for two excitation
light polarizations on a frequency scale. \ This peak shifting is the
phenomena that the diode-laser-stabilization schemes \cite{yas1,cli1,cor1}
mentioned in the introduction exploit.

As we can see from Figs.\ref{exp1}, \ref{exp2}, \ref{exp3}, \ref{exp4}, the
transitions between individual magnetic sublevels $m_{Fg}$ $\rightarrow$
$m_{Fe}$ of the same hyperfine transition $F_{g}$ $\rightarrow$ $F_{e}$ are
not resolved in the conditions of present experiment. Indeed, in the case of
complete resolution, one would expect to up to $9$ transitions between
different magnetic sublevels in case of circularly polarized excitation for
$F_{i}=1\rightarrow F_{e}=0,1,2,(3)$ (cases ($b$) and ($c$) Fig. \ref{exp3})
and up to $22$ magnetic transitions for case of linearly polarized excitation
for $F_{i}=3\rightarrow F_{e}=(1),2,3,4$ (case ($a$) Fig. \ref{exp2}). The
results of this experiment can be rather interpreted as $B$-field dependent
frequency shift and variation of probabilities of the $\left\vert F_{g}%
,m_{g}\right\rangle $ $\rightarrow$ $\left\vert F_{e},m_{e}\right\rangle $
hyperfine transitions. In the present experiment an essential limitation for
the resolution comes from the width of excitation laser line which was equal
to $25$ MHz. This allows to hope that in future resolution of sub-Doppler
spectroscopy in ETC can be increased and can be close to the homogeneous line
width of atomic transitions.

\section{Laser-induced-fluorescence excitation spectra in a magnetic field.
Signal simulation}

As we see from the observed signals, there exists a well pronounced structure
in laser-induced-fluorescence excitation spectra. This structure changes when
the magnetic field is applied. It can not be observed in Doppler broadened
spectra. Similar structure was observed earlier by \cite{tre1} but only in an
extremely strong magnetic field with strength of the order of $1000$ G and
stronger. At this field strength electronic angular momentum $\mathbf{J}$ of
an alkali atom is decoupled from the nuclei spin angular momentum $\mathbf{I}$
and both angular momenta interact with external field practically
independently. In present study, $\mathbf{J}$ and $\mathbf{I}$ are only
partially decoupled, and we can observe Pashen -- Back\textsc{\ }effect.

To simulate fluorescence spectra of Rb atoms in a magnetic field of
intermediate strength we will use the following model.\ We will assume that
laser radiation is weak and absorption rate $\Gamma_{p}$ is small in
comparison with the relaxation rates in ground and excited states, denoted by
$\gamma$ and $\Gamma$ respectively; $\Gamma_{p}<\gamma,$ $\Gamma$. This
assumption is well justified by the observation that
laser-induced-fluorescence excitation spectrum is independent of the laser intensity.

The Hamilton operator of the atom in a magnetic field can be written as%

\begin{equation}
\widehat{H}=\widehat{H}_{0}+\widehat{H}_{HFS}-\mathbf{\mu}_{J}\mathbf{\cdot
B-\mu}_{I}\mathbf{\cdot B.} \label{eq7}%
\end{equation}
where $\widehat{H}_{0}$ is a Hamiltonian operator of unperturbed atom without
taking into account nuclei spin, $\widehat{H}_{HFS}$ represents hyperfine
interaction. The remaining two terms represent interaction of the electronic
magnetic moment $\mathbf{\mu}_{J}$ of atom and the nucleus magnetic moment
$\mathbf{\mu}_{I}$ with the external magnetic field $\mathbf{B}$. These
magnetic moments are connected with the respective electronic and spin angular
moments\textbf{\ }$\mathbf{J}$ and $\mathbf{I}$ of the atom
\begin{equation}
\mathbf{\mu}_{J}\mathbf{=-}\frac{g_{J}\mu_{B}}{\hbar}\mathbf{J,\qquad\mu}%
_{I}\mathbf{=-}\frac{g_{I}\mu_{0}}{\hbar}\mathbf{I,} \label{eq8}%
\end{equation}
where $\mu_{B}$ and $\mu_{0}$ are the Bohr and nuclear magnetons respectively
and $g_{J}$, $g_{I}$ are electrotonic and nuclear Lande factors. The action of
the magnetic field on the atom has two closely related effects. First,
magnetic sublevels of the hyperfine levels are mixed by the magnetic field:
\begin{equation}
\left\vert \gamma_{k}m\right\rangle =\sum_{F_{e}=J_{e}-I}^{F_{e}=J_{e}%
+I}C_{kF_{e}}^{\left(  e\right)  }(B,m)\left\vert F_{e},m\right\rangle
,\qquad\left\vert \eta_{j}\mu\right\rangle =\sum_{F_{g}=J_{g}-I}^{F_{g}%
=J_{g}+I}C_{jF_{g}}^{\left(  g\right)  }(B,\mu)\left\vert F_{g},\mu
\right\rangle , \label{eq9}%
\end{equation}
where $C_{kF_{e}}^{\left(  e\right)  }(B,m)$ and $C_{jF_{g}}^{\left(
g\right)  }(B,\mu)$ are mixing coefficients that depend on the field strength
and magnetic quantum number $m$ or $\mu$. The second effect is deviation of
the Zeeman magnetic sublevel splitting in the magnetic field for each
hyperfine level from the linear one. It means that the additional energy of
the magnetic sublevel obtained in the magnetic field is not any more linearly
proportional to the field strength. New atomic states $\left\vert \gamma
_{k}m\right\rangle $ and $\left\vert \eta_{j}\mu\right\rangle $ in the
magnetic field in a general case are a linear combination of all initial
hyperfine levels ($4$ in case of Rb atoms in the $5P_{3/2}$ state and $2$ in
case of Rb atom in the $5S_{1/2}$ state). As it is seen from Eq. (\ref{eq9}),
in the magnetic field hyperfine angular momentum quantum number $F$ ceases to
be a good quantum number, but magnetic quantum numbers $m$ and $\mu$ are still
good quantum numbers. This reflects the symmetry of the perturbation imposed
by the magnetic field and means that only hyperfine sublevels with the same
magnetic quantum numbers are mixed by the magnetic field.

The mixing coefficients $C_{kF_{e}}^{\left(  e\right)  }(B,m)$ and $C_{jF_{g}%
}^{\left(  g\right)  }(B,\mu)$ of the hyperfine states in the magnetic field
and energies of these levels in the field $^{\gamma_{k}}E_{m} $, $^{\eta_{j}%
}E_{\mu}$ can be found as eigenvectors and eigenvalues of the Hamilton matrix
(\ref{eq7}). In Figures \ref{Rb85energy} and \ref{Rb87energy} the energy
levels obtained by the Hamilton matrix diagonalization for two Rb atom
isotopes in the excited $5P_{3/2}$ state in the magnetic field are presented.

For Rb atoms in the ground state hyperfine level splitting is larger than in
an excited state. It is around $3$ or $7$ GHz for two $^{85}$Rb and $^{87}$Rb
isotopes respectively, see Fig. \ref{spectra_total}, which in both cases is
large in comparison to the magnetic sublevel energies obtained in the magnetic
field of $50$ G. As a result ground state energy levels in the magnetic field
can, to a very good level of approximation, be represented by the linear
Zeeman effect. Namely, $^{\eta_{j}}E_{\mu}=g_{\eta_{j}}\mu_{B}B\mu/\hbar$,
where $g_{\eta_{j}}$ is the Lande factor of the respective hyperfine level.
For very weakly mixed levels it still can be represented with the hyperfine
quantum number $F_{g}$. For $^{85}$Rb atoms in $5S_{1/2}$ state we have
$g_{\eta_{j}}=-1/3$ for $F_{g}=2$ and $g_{\eta_{j}}=1/3$ for $F_{g}=3$, and
for $^{87}$Rb in $5S_{1/2}$ state we have $g_{\eta_{j}}=-1/2$ for $F_{g}=1$
and $g_{\eta_{j}}=1/2$ for $F_{g}=2.$ In case of mixing of only two hyperfine
levels the Breit-Rabi formula can be used to find both mixing coefficients and
level energies (see for example \cite{alex1,sva1}.)

In the magnetic field excited state density matrix created by the laser light
can be written as (see for example \cite{auz4}.)%

\begin{equation}
^{kl}f_{mm^{\prime}}=\frac{\widetilde{\Gamma}_{p}}{\Gamma+i^{kl}\Delta
\omega_{mm^{\prime}}}\sum_{j\mu}\left\langle \gamma_{k}m\right\vert
\widehat{\mathbf{E}}_{exc}^{\ast}\mathbf{\cdot}\widehat{\mathbf{D}}\left\vert
\eta_{j}\mu\right\rangle \left\langle \gamma_{l}m^{\prime}\right\vert
\widehat{\mathbf{E}}_{exc}^{\ast}\mathbf{\cdot}\widehat{\mathbf{D}}\left\vert
\eta_{j}\mu\right\rangle ^{\ast}. \label{eq10}%
\end{equation}
and $^{kl}\Delta\omega_{mm^{\prime}}=(^{\gamma_{k}}E_{m}-^{\gamma_{l}%
}E_{m^{\prime}})/\hbar$ is the energy splitting of the magnetic sublevels $m$
and $m^{\prime}$ belonging to the excited state levels $k$ and $l$. Magnetic
quantum numbers of the ground state level $\eta_{j}$ are denoted by $\mu$ and
magnetic quantum numbers of the excited state level $\gamma_{k,l}$ by $m$ and
$m^{\prime}$. In this last expression it is assumed that two magnetic
sublevels of the excited state, that initially belonged to two different
hyperfine levels at some specific magnetic field strength, can have the same
energy and can be excited simultaneously and coherently. It means nonzero
field level crossing signals \cite{auz4} in general can be included in this
model. At the same time for practical calculations performed in this work this
is not important, as far as laser radiation used to excite atoms in this work
is chosen in such a way that it can be represented only by one component in a
cyclic system of coordinates and is not able to create coherence between
magnetic sublevels of the atom.

In our particular simulation we assume that when we scan laser frequency only
those transitions that are in an exact resonance with the laser field are
excited.\ So at each laser frequency a specific density matrix is calculated.
This density matrix is of course dependent on the magnetic field strength that
determines magnetic sublevel splitting and wave function mixing. \ The
intensity of the fluorescence with \ a specific polarization $\mathbf{E}%
_{obs}$ in a transition between excited $\gamma_{k}$ and final $\eta_{j}$
state in the magnetic field can be calculated according to \cite{auz4}%

\begin{equation}
I\left(  \mathbf{E}_{f}\right)  =I_{0}\sum_{mm^{\prime}\mu}\sum_{klj}%
\left\langle \gamma_{k}m\right\vert \widehat{\mathbf{E}}_{obs}^{\ast
}\mathbf{\cdot}\widehat{\mathbf{D}}\left\vert \eta_{j}\mu\right\rangle
\left\langle \gamma_{l}m^{\prime}\right\vert \widehat{\mathbf{E}}_{obs}^{\ast
}\mathbf{\cdot}\widehat{\mathbf{D}}\left\vert \eta_{j}\mu\right\rangle ^{\ast
kl}f_{mm^{\prime}}. \label{eq4a}%
\end{equation}
The final state of the transition may or may not coincide with the atomic
hyperfine ground state level from which the absorption started. \ When in
expressions for excited state density matrix elements (Eq (\ref{eq10})) and
fluorescence intensity (Eq (\ref{eq4a})) ground $\left\vert \gamma
_{k}m\right\rangle $ and excited state $\left\vert \eta_{j}\mu\right\rangle $
wave function expansion over the atomic wave functions in absence of magnetic
field is used, see Eq. (\ref{eq9}) matrix elements of the type $\left\langle
F_{e}m\right\vert \widehat{\mathbf{E}}_{exc}^{\ast}\mathbf{\cdot}%
\widehat{\mathbf{d}}\left\vert F_{g}\mu\right\rangle $ appear. Standard
methods of angular momentum theory can be used to calculate these matrix
elements, see for example \cite{auz-mon,var2,sob1,zar1}.

\section{Analysis}

For laser-induced-fluorescence excitation spectra simulation in a magnetic
field we use the following procedure. From the magnetic sublevel energy
spectra obtained from the Hamilton matrix diagonalization we calculate
absorption line positions in a magnetic field. Then we calculate absorption
line strengths for these transitions taking into account level mixing, the
specific laser radiation polarization, and the magnetic field value. Then we
assume that the predominant factor in the formation of the absorption line
shape is homogeneous broadening. \ In the general case the profile is a Voigt
contour which is a convolution of homogeneous (Lorentzian) and inhomogeneous
(Gaussian) line shapes \cite{dem1}. Our assumption that inhomogeneous
broadening is substantially reduced in a sum-micron cell allows us to employ a
Lorentzian absorption profile. From the energy levels, intensities, and line
profiles, we are able to create spectra of expected fluorescence intensity in
the absence of spectral and polarization discrimination, but with specific
emission directions. It means that in this way we are simulating
laser-induced-fluorescence excitation spectra for our excitation --
observation conditions.

In Figures \ref{exp1}, \ref{exp2}, \ref{exp3}, \ref{exp4} calculated
intensities of fluorescence are shown as vertical bars of heights that
represent the intensity of fluorescence at a specific laser excitation
frequency. Solid lines in figures show laser-induced-fluorescence excitation
spectra that takes into account the linewidth of each absorption line as
described above. To get the best coincidence between measured and calculated
spectra the homogeneous linewidth of each absorption component was assumed to
be equal to $45$ MHz. This is several times the natural linewidth of the
resonant transition of unperturbed Rb atom. The radiation width of the
absorption components in D$_{2}$ line are expected to be approximately
$\Delta\nu_{D_{2}}=1/(2\pi\tau)\approx6$ MHz \cite{the1}, where $\tau
\approx26$ ns is lifetime of Rb in $5P_{3/2}$ state. Partially this absorption
line broadening is due to the fact that we have a broad laser line -- $25$
MHz. The rest of the broadening of the absorption line most probably can be
attributed to the atomic collisions within the cell walls.

\section{Conclusions}

For the first time it was demonstrated that ETC can be successfully used to
study spectra with sub-Doppler resolution of alkali atoms in a magnetic field.
In particular, for rather weak magnetic field of $50$ G when nucleus spin
angular momentum $\mathbf{I}$ and electronic angular momentum $\mathbf{J}$ are
only partially decoupled, we registered changes in laser-induced-fluorescence
excitation spectrum with sub-Doppler resolution of Rb atoms when different
polarizations of excitation laser are used. It was demonstrated that spectra
when right- and left-hand circularly polarized excitation is used differ very
strongly and these differences are not only quantitative, but also
qualitative. A strong magnetic field induced circular dichroism in this
experiment is clearly observed. Appearance of this circular dichroism may be
explained in a standard approach which describes how partial decoupling of
$\mathbf{I}$ and $\mathbf{J}$ mixes together magnetic sublevels of different
hyperfine states of alkali atoms. As a result we have a nonlinear Zeeman
effect for each magnetic sublevel. A second consequence of this mixing of wave
functions of different hyperfine states is a substantial modification of
transition probabilities between different magnetic sublevels, including the
fact that transitions that are strictly forbidden in absence of the magnetic
field start to be allowed. The importance of these changes of transition
probabilities even for the case of a rather weak magnetic field illustrates
that thin cells can show a sensitivity to physical processes not accessible to
conventional spectroscopy.

\section{Acknowledgments}

The authors are very thankful to Prof. Neil Shafer-Ray for fruitful
discussions. The authors are grateful to A. Sarkisyan for his valuable
participation in fabrication of the ETC. This work was supported, in part, by
ANSEF Grant \# PS 18-01 and Armenian Republic Grants \# 1351, 1323.

\section{Captions to the figures}%

%TCIMACRO{\FRAME{fhFUw}{2.7994in}{0.4454in}{0pt}{\Qcb{The design of the ETC.
%$1$ -- the windows of the ETC, made from a garnet (YAG) crystal, the thickness
%is $3$ mm, diameter is $30$ mm. The flatness of the both inner surfaces of the
%YAG windows is less than $\lambda/10$; $2$ -- $\sim$\ $600$ nm-thick Al$_{2}%
%$O$_{3}$\ coating in the shape of a $1$ mm-wide and $10$ mm-long strip; $3$ --
%the glue; $4$ -- the sapphire side arm with a \textquotedblright molybdenum
%glass\textquotedblright\ termination; $5$ -- column of Rb metal.}}{\Qlb{fig1}%
%}{Figure}{\special{ language "Scientific Word";  type "GRAPHIC";
%maintain-aspect-ratio TRUE;  display "ICON";  valid_file "T";
%width 2.7994in;  height 0.4454in;  depth 0pt;  original-width 2.757in;
%original-height 0.4151in;  cropleft "0";  croptop "1";  cropright "1";
%cropbottom "0";  tempfilename 'HCT5U600.wmf';tempfile-properties "PR";}}}%
%BeginExpansion
%EndExpansion%
%TCIMACRO{\FRAME{fhFUw}{2.7994in}{0.4454in}{0pt}{\Qcb{Excited-state hyperfine
%magnetic sublevel splitting for intermediate strength of a magnetic field for
%$^{85}$Rb isotope}}{\Qlb{Rb85energy}}{Figure}%
%{\special{ language "Scientific Word";  type "GRAPHIC";
%maintain-aspect-ratio TRUE;  display "ICON";  valid_file "T";
%width 2.7994in;  height 0.4454in;  depth 0pt;  original-width 2.757in;
%original-height 0.4151in;  cropleft "0";  croptop "1";  cropright "1";
%cropbottom "0";  tempfilename 'HCT5U601.wmf';tempfile-properties "PR";}}}%
%BeginExpansion
%EndExpansion%
%TCIMACRO{\FRAME{fhFUw}{2.7994in}{0.4454in}{0pt}{\Qcb{Excited-state hyperfine
%magnetic sublevel splitting for intermediate strength of a magnetic field for
%$^{87}$Rb isotope.}}{\Qlb{Rb87energy}}{Figure}%
%{\special{ language "Scientific Word";  type "GRAPHIC";
%maintain-aspect-ratio TRUE;  display "ICON";  valid_file "T";
%width 2.7994in;  height 0.4454in;  depth 0pt;  original-width 2.757in;
%original-height 0.4151in;  cropleft "0";  croptop "1";  cropright "1";
%cropbottom "0";  tempfilename 'HCT5U602.wmf';tempfile-properties "PR";}}}%
%BeginExpansion
%EndExpansion%
%TCIMACRO{\FRAME{fhFUw}{2.7994in}{0.4454in}{0pt}{\Qcb{Doppler broadened
%fluorescence spectra (light line) obtained in an ordinary cell and sub-Doppler
%spectra obtained in ETC on D$_{2}$ line of both Rb isotopes in absence of the
%magnetic field.}}{\Qlb{spectra_total}}{Figure}%
%{\special{ language "Scientific Word";  type "GRAPHIC";
%maintain-aspect-ratio TRUE;  display "ICON";  valid_file "T";
%width 2.7994in;  height 0.4454in;  depth 0pt;  original-width 2.757in;
%original-height 0.4151in;  cropleft "0";  croptop "1";  cropright "1";
%cropbottom "0";  tempfilename 'HCT5U603.wmf';tempfile-properties "PR";}}}%
%BeginExpansion
%EndExpansion%
%TCIMACRO{\FRAME{fhFUw}{2.7994in}{0.4454in}{0pt}{\Qcb{Geometrical configuration
%of the experiment. Magnetic field is applied along $z$ or $x$ axis. }%
%}{\Qlb{geom}}{Figure}{\special{ language "Scientific Word";  type "GRAPHIC";
%maintain-aspect-ratio TRUE;  display "ICON";  valid_file "T";
%width 2.7994in;  height 0.4454in;  depth 0pt;  original-width 2.757in;
%original-height 0.4151in;  cropleft "0";  croptop "1";  cropright "1";
%cropbottom "0";  tempfilename 'HCT5U604.wmf';tempfile-properties "PR";}}}%
%BeginExpansion
%EndExpansion%
%TCIMACRO{\FRAME{fhFUw}{2.7994in}{0.4454in}{0pt}{\Qcb{Experimental (scatter
%circle) and simulated (dark line) fluorescence spectra of $^{85}$Rb isotope on
%transitions $F_{g}=2\rightarrow F_{e}=1,2,3,(4)$. ($a$) linearly polarized
%$\pi$ excitation, ($b$) circularly polarized $\sigma^{-}$ excitation, ($c$)
%circularly polarized $\sigma^{+}$ excitation. With light line respective
%fluorescence spectra in absence of the magnetic field is shown.}}{\Qlb{exp1}%
%}{Figure}{\special{ language "Scientific Word";  type "GRAPHIC";
%maintain-aspect-ratio TRUE;  display "ICON";  valid_file "T";
%width 2.7994in;  height 0.4454in;  depth 0pt;  original-width 2.757in;
%original-height 0.4151in;  cropleft "-0.159729";  croptop "1";
%cropright "0.840271";  cropbottom "0";
%tempfilename 'HCT5U605.wmf';tempfile-properties "PR";}}}%
%BeginExpansion
%EndExpansion%
%TCIMACRO{\FRAME{fhFUw}{2.7994in}{0.4454in}{0pt}{\Qcb{Experimental (scatter
%circle) and simulated (dark line) fluorescence spectra of $^{85}$Rb isotope on
%transitions $F_{g}=3\rightarrow F_{e}=(1),2,3,4$. ($a$) linearly polarized
%$\pi$ excitation, ($b$) circularly polarized $\sigma^{-}$ excitation, ($c$)
%circularly polarized $\sigma^{+}$ excitation. With light line respective
%fluorescence spectra in absence of the magnetic field is shown.}}{\Qlb{exp2}%
%}{Figure}{\special{ language "Scientific Word";  type "GRAPHIC";
%maintain-aspect-ratio TRUE;  display "ICON";  valid_file "T";
%width 2.7994in;  height 0.4454in;  depth 0pt;  original-width 2.757in;
%original-height 0.4151in;  cropleft "0";  croptop "1";  cropright "1";
%cropbottom "0";  tempfilename 'HCT5U606.wmf';tempfile-properties "PR";}}}%
%BeginExpansion
%EndExpansion%
%TCIMACRO{\FRAME{fhFUw}{2.7994in}{0.4454in}{0pt}{\Qcb{Experimental (scatter
%circle) and simulated (dark line) fluorescence spectra of $^{87}$Rb isotope on
%transitions $F_{g}=1\rightarrow F_{e}=0,1,2,(3)$. ($a$) linearly polarized
%$\pi$ excitation, ($b$) circularly polarized $\sigma^{-}$ excitation, ($c$)
%circularly polarized $\sigma^{+}$ excitation. With light line respective
%fluorescence spectra in absence of the magnetic field is shown.}}{\Qlb{exp3}%
%}{Figure}{\special{ language "Scientific Word";  type "GRAPHIC";
%maintain-aspect-ratio TRUE;  display "ICON";  valid_file "T";
%width 2.7994in;  height 0.4454in;  depth 0pt;  original-width 2.757in;
%original-height 0.4151in;  cropleft "0";  croptop "1";  cropright "1";
%cropbottom "0";  tempfilename 'HCT5U607.wmf';tempfile-properties "PR";}}}%
%BeginExpansion
%EndExpansion%
%TCIMACRO{\FRAME{fhFUw}{2.7994in}{0.4454in}{0pt}{\Qcb{Experimental (scatter
%circle) and simulated (dark line) fluorescence spectra of $^{87}$Rb isotope on
%transitions $F_{g}=2\rightarrow F_{e}=1,2,3,(4)$. ($a$) linearly polarized
%$\pi$ excitation, ($b$) circularly polarized $\sigma^{-}$ excitation, ($c$)
%circularly polarized $\sigma^{+}$ excitation. With light line respective
%fluorescence spectra in absence of the magnetic field is shown.}}{\Qlb{exp4}%
%}{Figure}{\special{ language "Scientific Word";  type "GRAPHIC";
%maintain-aspect-ratio TRUE;  display "ICON";  valid_file "T";
%width 2.7994in;  height 0.4454in;  depth 0pt;  original-width 2.757in;
%original-height 0.4151in;  cropleft "0";  croptop "1";  cropright "1";
%cropbottom "0";  tempfilename 'HCT5U608.wmf';tempfile-properties "PR";}}}%
%BeginExpansion
%EndExpansion%
%TCIMACRO{\FRAME{fhFUw}{2.7994in}{0.4454in}{0pt}{\Qcb{Absorption spectra in an
%ordinary optical cell for ($a$) $^{85}$Rb and ($b$) $^{87}$Rb. Peak 1 is
%absorption for $\sigma^{-}$ or $\sigma^{+}$ radiation in absence of the
%magnetic field. Peak 2 corresponds to the absorption of $\sigma^{-}$ radiation
%in a magnetic field and peak 3 to the absorption of $\sigma^{+}$ radiation.}%
%}{\Qlb{exp_broad}}{Figure}{\special{ language "Scientific Word";
%type "GRAPHIC";  maintain-aspect-ratio TRUE;  display "ICON";
%valid_file "T";  width 2.7994in;  height 0.4454in;  depth 0pt;
%original-width 2.757in;  original-height 0.4151in;  cropleft "0";
%croptop "1";  cropright "1";  cropbottom "0";
%tempfilename 'HCT5U609.wmf';tempfile-properties "PR";}}}%
%BeginExpansion
%EndExpansion


\begin{thebibliography}{99}                                                                                               %


\bibitem {sch1}A.L. Schawlow, Rev. Mod. Phys. 54(3) (1982) 697

\bibitem {let1}V. S. Letokhov, V.P. Chebotayev, Nonlinear Laser Spectroscopy,
Springer Series in Optical Sciences, Vol. 4, Springer, Berlin, 1977

\bibitem {wyn1}R. Wynands, A. Nagel, Appl. Phys. B 68 (1999) 1

\bibitem {gla1}D.S. Glassner, B. Ai, R.J. Knize, Opt. Lett. 19 (1994) 2071

\bibitem {ai1}B. Ai, D.S. Glassner, R.J. Knize, J.P. Partanen, Appl. Phys.
Lett. 64(8) (1994) 951

\bibitem {bri1}S. Briaudeau, D. Bloch, M. Ducloy, Europhys. Lett. 35(5) (1996) 337

\bibitem {bri2}S. Briaudeau, S. Saltiel, G. Nienhuis, D. Bloch, M. Ducloy,
Phys. Rev. A 57 (1998) R3169

\bibitem {bri3}S. Briaudeau, D. Bloch, M. Ducloy, Phys. Rev. A 59 (1999) 3723

\bibitem {bri4}S. Briaudeau, S. Saltiel, J.R.R. Leite, M. Oria, A. Bramati, A.
Weis, D. Bloch, M. Ducloy, J. Phys. IV France 10 (2000) Pr8-145

\bibitem {izm1}A. Izmailov, Laser Phys. 3 (1993) 507

\bibitem {rom1}R.H. Romer, R.H. Dicke, Phys. Rev. 99 (1955) 532

\bibitem {var1}T.A. Vartanyan, D.L. Lin, Phys. Rev. A 38 (1995) 5197

\bibitem {zam1}B. Zambon, G. Nienhuis, Opt. Commun. 143 (1997) 308

\bibitem {yar1}A. Yarovitski, G. Dutier, S. Saltiel et al. , Int. Quant.
Electr. Conf. IQEC/2002, Technical Digest, p. 333, Moscow, Russia, 2002

\bibitem {dut1}G. Dutier, S. Saltiel, D. Bloch, M. Ducloy, \textit{to be
published}

\bibitem {sar2}D. Sarkisyan, D. Bloch, A. Papoyan, M. Ducloy, Opt. Commun. 200
(2001), 201

\bibitem {khr1}I.B. Khriplovich, S.K. Lamoreaux, \textit{CP Violation Withour
Strangeness, Electric Dipole Moments of Particles, Atoms, and Molecules},
Springer, Berlin, Heidelberg, New York, Barcelona, Budapest, Hong Kong,
London, Milan, Paris, Santa Clara, Singapore, Tokyo (1997) 230

\bibitem {com1}E.D. Commins, Physica Scripta 46 (1993) 92

\bibitem {bou1}M.A. Bouchiat, C. Bouchiat, Rep. Progr. Phys. 60 (1997) 1351

\bibitem {gue1}J. Guena, D. Chauvat, Ph. Jacquier, E. Jahier, M. Lintz, S.
Sanguinetti, A. Wasan, M.A. Bouchiat, A.V. Papoyan, D. Sarkisyan, Phys. Rev.
Lett., 90 (2003), 143001

\bibitem {yas1}V.V. Yashchuk, D. Budker, J.R. Davis, Rev. Sci. Instrum. 71(2)
(2000) 341

\bibitem {cli1}M.A. Clifford, G.P.T. Lancaster, R.S. Conroy, K. Dholakia, J.
Mod. Opt. 47(11) (2000) 1933

\bibitem {cor1}K.L. Corwin, Z.T. Lu, C.F. Hand, R.J. Epstein, C.E. Wieman,
Appl. Opt. 37(15) (1998) 3295

\bibitem {nov1}V.N. Novikov, O.P. Sushkov, I.B. Khriplovich, Opt. Spectrosc.
42 (1977) 370; Opt. Spectrosc. 45 (1978) 236

\bibitem {rob1}G.J. Roberts, P.E.G. Baird, M.W.S. Brimicombe, P.G.H. Sandars,
D.R. Selby, D.N. Stacey, J. Phys. B: Atom. Molec. Phys. 13 (1980) 1389

\bibitem {che1}X. Chen, V.L. Telegdi, A. Weis, J. Phys. B: Atom. Molec. Phys.
20 (1987) 5653

\bibitem {tre1}P. Tremblay, A. Michaud, M. Levesque, S. Theriault, M. Breton,
J. Beaubien, N. Cyr, Phys. Rev. A 42(5) (1990) 2766

\bibitem {wei1}A. Weis, V.A. Sautenkov, T.W. Hansch, J. Phys. II France 3
(1993) 263

\bibitem {pap1}N. Papageorgiou, A. Weis, V.A. Sautenkov, D. Bloch, M. Ducloy,
Appl. Phys. B 59 (1994) 123

\bibitem {the1}C.E. Theodosiou, Phys. Rev. A 30 (1984) 2881

\bibitem {auz-mon}M. Auzinsh, R. Ferber, \textit{Optical Polarization of
Molecules, }Cambridge University Press, Cambridge (1995)

\bibitem {zar1}R.N. Zare, \textit{Angular Momentum,} J. Wiley and Sons, New
York (1988)

\bibitem {var2}D.A. Varshalovich, A.N. Moskalev, V.K.
Khersonskii,\textit{\ Quantum Theory of Angular Momentum}, World Scientific,
Singapore (1988)

\bibitem {sob1}I.I. Sobelman, \textit{Atomic Spectra and Radiative
Transitions}, Springer Verlag, Berlin (1979)

\bibitem {alex1}E.B. Aleksandrov, M.P. Chaika, G.I. Khvostenko,
\textit{Interference of Atomic States}, Springer Verlag, \ New York, (1993)

\bibitem {sva1}S. Svanberg, \textit{Atomic and Molecular Spectroscopy: Basic
Aspects and Prectical Applications}, Springer, Berlin, Heidelberg (1992)

\bibitem {auz4}J. Alnis, M. Auzinsh, Phys. Rev. A\ 63 (2001) 023407

\bibitem {dem1}W. Demtroder, \textit{Laser spectroscopy: Basic Concepts and
Instrumentation,} Springer, Berlin, Heidelberg, New York, Hong Kong, London,
Milan, Paris, Tokyo (2002)
\end{thebibliography}
\end{document}